\documentclass[aps,preprint,showpacs]{revtex4}
\begin{document}
\title{Unitary transformations purely based on geometric phases \\
of two spins with Ising interaction in a rotating magnetic field}
\author{Yu Shi}
\thanks{Email: yushi@fudan.edu.cn}
\affiliation{Department of Physics, Fudan University, Shanghai
200433, China}

\begin{abstract}
We consider how to obtain a nontrivial two-qubit unitary
transformation purely based on geometric phases of two
spin-$\frac{1}{2}$'s with Ising-like interaction in a magnetic field
with a static $z$-component and a rotating $xy$-component. This is
an interesting problem both for the purpose of measuring the
geometric phases and in quantum computing applications. In previous
approach, coupling of one of the qubit with the rotating component
of field is ignored. By considering the exact two-spin geometric
phases, we find that a nontrivial two-spin unitary transformation
purely based on Berry phases can be obtained by using two
consecutive cycles with opposite directions of the magnetic field
and opposite signs of the interaction constant. In the nonadiabatic
case, starting with a certain initial state, a cycle in the
projected space of rays and thus Aharonov-Anandan phase can be
achieved. The two-cycle scheme cancels the total phases, hence any
unknown initial state evolves back to itself without a phase factor.
\end{abstract}

\pacs{03.65.Vf, 03.67.Lx} \maketitle

\section{Introduction}

In some circumstances, when a Hamiltonian undergoes a cyclic
evolution, the quantum state picks up a geometric phase in addition
to the dynamical phase. The adiabatic geometric phase, called Berry
phase, is accumulated in an instantaneous eigenstate of an
adiabatically varying Hamiltonian that is cyclic in the parameter
space~\cite{berry}. The nonadiabatic geometric phase, called
Aharonov-Anandan phase, is the geometric part of the phase
accumulated in a state which is cyclic in the projected space of
rays~\cite{aharonov}. The last two decades witnessed many
fascinating applications of geometric phases in many areas of
physics~\cite{shapere}, most of which use geometric phases of single
particles or spins. In recent years, new perspectives were opened in
considering their possible applications in quantum computing,
motivated by their insensitivity to dynamical
details~\cite{sjoeqvist}.  This so-called geometric quantum
computing shares some features with, though not as robust as,
topological quantum computing~\cite{topo}. The uses of both the
usual Abelian geometric phases~\cite{condition,jones,zhu} and the
non-Abelian geometric phases~\cite{zanardi} were proposed.
Nonadiabatic error in quantum computing based on Berry phases was
analyzed~\cite{shi2}. One approach to realizing geometric quantum
gates uses optical control in the computational Hilbert space
together with ancillary states~\cite{duan}.

Here we examine another approach, namely, the direct use of an
interacting two-qubit Hamiltonian to construct nontrivial two-qubit
unitary transformation purely based on geometric phases. In this
approach, Hamiltonian involves only the computational Hilbert space,
without ancillary space. For isotropic Heisenberg interaction,
however, it is impossible to construct a nontrivial two-qubit
transformation purely based on Berry phases, or purely based on
Aharonov-Anandan phases if the two qubits have equal gyromagnetic
ratios, as the Hamiltonian is too symmetric~\cite{shi1}. Another
form of the interaction that often appears in quantum computing
implementations is the Ising interaction, which is an approximation
to the isotropic Heisenberg interaction in the liquid
NMR~\cite{nielsen}, and can also be realized in cold atoms in
optical lattices~\cite{jaksch},  Josephson junction
circuits~\cite{schoen}, and so on. Ising interaction was used in a
pioneering construction of conditional phase gate based on geometric
phases~\cite{condition}, which pointed out that a crucial issue is
to cancel the dynamical phases. Nevertheless, as explained in the
next section, the treatment, followed by some other papers, is
unsatisfactory.

In this paper, we reexamine the general situation of two
Ising-interacting spin-$\frac{1}{2}$'s equally coupled with a
rotating magnetic field, and construct a unitary transformation
purely based on the geometric phases of the two-body system, rather
than reduce it to geometric phases of a single spin. For adiabatic
case, we discuss a two-cycle scheme using two cycles with opposite
directions of the field and opposite signs of the Ising interaction
constant. At the end of these two cycles, the dynamical phases are
always canceled, so that the phases become purely geometric.
Consequently, one obtains a unitary transformation purely based on
Berry phases, which can act on any unknown initial state. For
Aharonov-Anandan phases, the analogous two-cycle scheme yields a
unity transformation, i.e. any unknown initial state evolves back
without a phase factor.

The rest of the paper is organized as follows. The Hamiltonian and
the problem are introduced in Sec.~II. The adiabatic evolution is
discussed in Sec.~III. In this section, after the calculation of
Berry phases in instantaneous eigenstates in Subsec.~A, a unitary
transformation purely based on Berry phases is constructed by using
a two-cycle scheme in Subsec.~B. In Sec.~IV, we study the exact
evolution without the adiabatic condition. Following the exact
solution of the Sch\"{o}dinger equation in Subsec.~A, the cycling
and Aharonov-Anandan phases are discussed in Subsec.~B, for initial
states which are prepared to be eigenstates of a certain
time-independent Hamiltonian. The two-cycle scheme for
Aharonov-Anandan phases is discussed in Subsec.~C. Summary and
discussions are  given in Sec.~V. We set $\hbar=1$ throughout this
article.

\section{The Hamiltonian}

The magnetic field $\mathbf{B}$ has a static component of magnitude
$B_0$ along $z$ direction and a rotating component of magnitude
$B_1$  on $xy$ plane. It can be written as
$$\mathbf{B}=B_1\cos\omega_1t \hat{\mathbf{x}}
+B_1\sin\omega_1t \hat{\mathbf{y}}+B_0\hat{\mathbf{z}},$$ where
$\hat{\mathbf{x}}$, $\hat{\mathbf{y}}$ and $\hat{\mathbf{z}}$ are
unit vectors along these three directions, $\omega_1$ is the angular
velocity of the rotation of the $xy$-component of the field.

For two Ising-interacting spin-$\frac{1}{2}$'s in this magnetic
field, the Hamiltonian can be written as
\begin{equation}
{\cal H} (t) = {\cal H}_0+{\cal H}_{a1}+{\cal H}_{b1}, \label{ham}
\end{equation} with
\begin{equation} {\cal H}_0 = \frac{1}{2}(\omega_{a0}\sigma_{az} + \omega_{b0}\sigma_{bz} +
J\sigma_{az}\sigma_{bz}), \end{equation} \begin{equation} {\cal
H}_{\alpha 1} = \frac{\Gamma_{\alpha }}{2}(\sigma_{\alpha x}\cos
\omega_{1}t + \sigma_{\alpha y} \sin \omega_{1} t). \end{equation}
${\cal H}_0$, which is time-independent, includes the coupling of
the two spins with the static $z$-component of the field, as well as
the Ising interaction between them. ${\cal H}_{\alpha 1}$, which is
time-dependent, is the coupling of spin $\alpha$ ($\alpha=a,b$) with
the rotating $xy$-component of the field. $\omega_{\alpha
0}=-\kappa_{\alpha} B_0$, $\Gamma_{\alpha}=-\kappa_{\alpha} B_1$,
$\kappa_{\alpha}$¡¡is the gyromagnetic ratio of spin $\alpha$.  The
variation period of the Hamiltonian is $\tau= 2\pi/\omega_1$.

Previous approach is the following~\cite{condition}. Starting with a
basis state $|\sigma_{az}\sigma_{bz}\rangle$, in which each spin is
along $z$ direction, $\sigma_{az}$ is regarded as conserved, while
$|\sigma_{bz}\rangle$ is transformed by a geometric phase
conditional on the value of $\sigma_{az}$.  In this approach, the
Hamiltonian is actually regarded as block diagonal with respect to
$\sigma_{az}$, i.e. ${\cal H}_{a1}$ is completely ignored. In fact,
now that both spins are coupled with the rotating $xy$-component of
the field,  spin $a$ cannot remain as an eigenstate of the operator
$\sigma_{az}$. Moreover,  Berry phase should be accumulated in an
eigenstate of the Hamiltonian, rather than in
$|\sigma_{az}\sigma_{bz}\rangle$, which is not an eigenstate of the
Hamiltonian no matter whether ${\cal H}_{a1}$ is negligible.
Consequently the unitary transformation based on geometric phases is
not diagonal in basis \{$|\sigma_{az}\sigma_{bz}\rangle$\}, as
described in Ref.~\cite{condition}, which is problematic even if
${\cal H}_{a1}=0$. If ${\cal H}_{a1}=0$, the eigenstate of ${\cal
H}$ at $t=0$ is $|\sigma_{az}\sigma_{b\mathbf{n}}\rangle$, where
$\mathbf{n}$ is the direction along
$(\omega_{b0}+J\sigma_{az})\hat{\mathbf{z}}+\Gamma_{b}\hat{\mathbf{x}}$.
Berry phase  accumulates in
$|\sigma_{az}\sigma_{b\mathbf{n}}\rangle$, instead of
$|\sigma_{az}\sigma_{bz}\rangle$, and  can be obtained as
$-\pi[1-\cos\Theta(\sigma_{az})]$, where $\Theta(\sigma_{az})=
\tan[(\omega_{b0}+J\sigma_{az})/\Gamma_{b}$. In some other
papers~\cite{jones,zhu}, the coupling of one of the qubits with the
rotating component of the field was noticed and neglected on the
condition that it is far off resonance.

In the following, we shall focus on the case that the gyromagnetic
ratios of the two spins are the same, i.e.
$\omega_{a0}=\omega_{b0}=\omega_0$, $\Gamma_a=\Gamma_b=\Gamma$. In
this case, the two spins are {\em equally} coupled with the rotating
magnetic field, hence none of them can ever be approximated as
decoupled with the rotating field. Moreover, that the two qubits are
identical is also convenient for scalable quantum computing.

\section{Berry Phases}

\subsection{Berry phases in instantaneous eigenstates}

Let us first consider the adiabatic limit of the Hamiltonian, and
give its instantaneous eigenstates, written in terms of the
eigenstates $|\sigma_{az}\sigma_{bz}\rangle$ of  the $z$-component
spin operators as
$$|\xi_n(t)\rangle \equiv x_n|\uparrow\uparrow\rangle
+y_n|\uparrow \downarrow\rangle+z_n|\downarrow
\uparrow\rangle+w_n|\downarrow \downarrow\rangle,$$ where
$n=1,2,3,4$.

One of the eigenstates is just the singlet state
\begin{equation}
|\xi_4\rangle = \frac{1}{\sqrt{2}}(|\uparrow\downarrow\rangle-
|\downarrow\uparrow\rangle),
\end{equation}
with eigenvalue
\begin{equation}
E_4=-J/2.
\end{equation}
This can be understood by noting that the Hamiltonian
can be written in terms of the total spin as ${\cal H}= \omega_0 S_z
+J S^2_z+\Gamma(S_x\cos \omega_1t +S_y \sin \omega_1 t)-J/2$,
consequently the singlet state with total spin $S=0$ must be its
eigenstate.

The other three instantaneous eigenstates $|\xi_n(t)$'s ($n=1,2,3$)
are found to be
\begin{widetext}
\begin{equation}
|\xi_n\rangle=\frac{1}{\sqrt{{\cal N}_n}} (\frac{-2\Gamma
e^{-i\omega
t}}{2\omega_{0}+J-2E_n}|\uparrow\uparrow\rangle+|\uparrow
\downarrow\rangle+|\downarrow \uparrow\rangle+\frac{2\Gamma
e^{i\omega t}}{-2\omega_{0}+J-2E_n}|\downarrow \downarrow\rangle),
\label{xin}
\end{equation}
\end{widetext}  with  \begin{equation}{\cal N}_n= 2+
\frac{4\Gamma^2}{(2\omega_0+J-2E_n)^2}+
\frac{4\Gamma^2}{(-2\omega_0+J-2E_n)^2}.
\end{equation} The corresponding
eigenvalues,  satisfying
$(2E_n)^3-J(2E_n)^2-(J^2+4\omega_0^2+4\Gamma^2)(2E_n)+
(J^3-4\omega_0^2J+4\Gamma^2J)=0$, are found to be~\cite{math}
\begin{equation} E_n=\frac{1}{2}(v_n+\frac{J}{3}),
\end{equation} where $v_n$'s are solutions of
the equation
\begin{equation}v^3+pv+q=0,
\end{equation} with
\begin{equation} \begin{array}{rcl}
p& = & -(\frac{4J^2}{3} + 4\omega_0^2 +4\Gamma^2),\\
q& = & \frac{16}{27}J^3 +\frac{(8\Gamma^2-16\omega_0^2)J}{3}.
\end{array} \label{pq}   \end{equation}
As $p<0$, we obtain
\begin{equation} \begin{array}{rcl}
E_1&=&\sqrt{-\frac{p}{3}} \cos\Phi + \frac{J}{6}, \\
E_2&=&\sqrt{-\frac{p}{3}} \cos (\Phi+\frac{2\pi}{3}) + \frac{J}{6}
,\\
E_3&=&\sqrt{-\frac{p}{3}} \cos (\Phi-\frac{2\pi}{3})+
\frac{J}{6},\end{array} \label{e}   \end{equation} where
\begin{equation}\Phi= \frac{1}{3} \arccos
(-\frac{q}{2\sqrt{-(\frac{p}{3})^3}}).\label{phi}   \end{equation}

In the adiabatic limit, each eigenstate evolves as
\begin{equation} |\xi_n(\tau)\rangle =
e^{i\gamma_n}|\xi_n(0)\rangle,\end{equation}  with the total phase
factor
\begin{equation}\gamma_n = \gamma_n^d+\gamma_n^B, \label{total}
\end{equation}
where
\begin{equation} \gamma_n^d = -E_n\tau\end{equation}  is the dynamical phase,
while \begin{equation} \gamma^B_n = \int_0^{\tau} dt \langle
\xi_n(t)|\partial_t \xi_n(t)\rangle\end{equation}  is the Berry
phase. It is found that \begin{equation} \gamma^B_n =
2\pi(|x_n|^2-|w_n|^2).\end{equation}  Therefore, it is obtained that
\begin{equation}
\gamma_n^B = \frac{32\Gamma^2\omega_0(2E_n-J)}{{\cal N}_n[(2E_n-
J)^2 -4 \omega_0^2]^2} \label{gammaa}
\end{equation}
for $n=1,2,3$, while \begin{equation} \gamma_4^B=0.\end{equation}

As an indication that $\gamma_n^B$ is geometric, $\omega_1$ does not
appear in the above expressions of $\gamma_n^B$'s. In fact,
$\omega_1 t$ in the Hamiltonian can be replaced as an angle variable
$\theta$, thus
$$\gamma^B_n = \int_0^{2\pi} dt \langle
\xi_n(\theta)|\partial_{\theta}
\xi_n(\theta)\rangle,$$ which is independent of how $\theta$
varies with time.

The instantaneous eigenstates and Berry phases have already been
calculated by Yi and Sj\"{o}qvist in a different form~\cite{yi}.

\subsection{Geometric transformation of an arbitrary unknown state}

For the purposes of both experimentally detecting geometric phases
and the application in quantum computing, it is crucial to cancel
the dynamical phases, as pioneered in Ref.~\cite{jones,condition}.
With the correct geometric and dynamical phases of this two-body
problem, instead of the incorrect reduction to a one-body problem,
it becomes more nontrivial to cancel the dynamical phases.

First we note that $E_n$ depends on $\omega_0^2$ and $\Gamma^2$,
while depends on $J$. When $J$  reverses its sign, $q$ reverses sign
while $p$ remains unchanged, according to Eq.~(\ref{pq}).
Consequently, Eq.~(\ref{e}) indicates that there exist the following
relations
\begin{equation} \begin{array}{rcl}
E_1[-J] & = & -E_2[J], \\
E_2[-J] & = & -E_1[J],  \\
E_3[-J] & = & -E_3[J], \\
E_4[-J] & = & -E_4[J]. \end{array}  \label{er} \end{equation}

Then according to expression of $|\xi_n\rangle$ in Eq.~(\ref{xin}),
if $\Gamma$, $\omega_0$ and $J$ all reverse signs, $|\xi_1\rangle$
and $|\xi_2\rangle$ exchange, while $|\xi_3\rangle$ and
$|\xi_4\rangle$ remain unchanged, i.e.,
\begin{equation} \begin{array}{rcl}
|\xi_1[-\Gamma,-\omega_{0},-J]\rangle &=&
|\xi_2[\Gamma,\omega_{0},J]\rangle, \\
|\xi_2[-\Gamma,-\omega_{0},-J]\rangle &=&
|\xi_1[\Gamma,\omega_{0},J]\rangle, \\
|\xi_3[-\Gamma,-\omega_{0},-J]\rangle & =&
\xi_3[\Gamma,\omega_{0},J]\rangle, \\
|\xi_4[-\Gamma,-\omega_{0},-J]\rangle & =&
\xi_4[\Gamma,\omega_{0},J]\rangle.\end{array} \label{sr}
\end{equation}

According to the expression of $\gamma^B_n$ in  Eq.~(\ref{gammaa}),
there exist the following relations
\begin{equation} \begin{array}{rcl}
\gamma_1^B [-\omega_{0},-J] &=&\gamma_2^B
[\omega_{0},J] \\
\gamma_2^B [-\omega_{0},-J] &=&\gamma_1^B
[\omega_{0},J] \\
\gamma_3^B [-\omega_{0},-J] &=&\gamma_3^B
[\omega_{0},J] \\
\gamma_4^B [-\omega_{0},-J] &=&\gamma_4^B [\omega_{0},J]
=0.\end{array} \label{berry}
\end{equation}
Note that $E_i$ and thus $\gamma_i^d$ as well as $\gamma_i^B$ depend
only on the magnitude of $\Gamma$, independent of its sign, as
implied by the fact that $xy$-component of the field rotates.

Now consider an arbitrary initial unknown state $|\Psi(0)\rangle$,
which evolves  under the adiabatically varying Hamiltonian.  After
$\tau$,  the state becomes
\begin{equation} |\Psi(\tau)\rangle = U |\Psi(0)\rangle, \end{equation}
where \begin{equation} U =  diag (e^{i\gamma_1}, e^{i\gamma_2},
e^{i\gamma_3}, e^{i\gamma_4}),\end{equation}  written in the basis
$\{|\xi_1\rangle, |\xi_2\rangle, |\xi_3\rangle, |\xi_4\rangle\}$,
where $\gamma_n$'s are the total phases as given in
Eq.~(\ref{total}).

In the following, we propose a two-cycle method to cancel the
dynamical phase $\gamma_n^d$ for all eigenstates simultaneously,
i.e. only the Berry phases $\gamma^B_n$'s appear in the final
unitary transformation, consequently,
\begin{equation} |\Psi(2\tau)\rangle = U_B |\Psi(0)\rangle,
\label{geo}\end{equation} where \begin{equation} U_B = diag
(e^{2i\gamma^B_1}, e^{2i\gamma^B_2}, e^{2i\gamma^B_3},
e^{2i\gamma^B_4}).\end{equation}  with $\gamma^B_n$'s  given in
Eq.~(\ref{gammaa}).

We make the Hamiltonian undergo two consecutive cycles, with the
signs of $J$, $\omega_{0}$ and $\Gamma$  all reversed in the second
cycle. The reversal of these signs must be done at the end of the
first cycle instantaneously. Because of the relations of eigenvalues
(\ref{er}) and of instantaneous eigenstates (\ref{sr}), under sign
reversal of  $J$, $\omega_{0}$ and $\Gamma$,  an instantaneous
eigenstate at the end of the first cycle continues to be an
eigenstate in the second cycle, but with the sign of the eigenvalue
reversed. Consequently, the dynamical phases in the two cycles
cancel.

To comprehend the situation clearly, suppose the state starts as an
instantaneous eigenstate $|\xi_1[\Gamma,\omega_{0},J]\rangle$. In
the first cycle of the Hamiltonian, it remains as the instantaneous
eigenstate, accumulating a phase
$\gamma_1^B[\omega_{0},J]-E_1[J]\tau$ at the end of the first cycle.
Then $\Gamma$, $\omega_{0}$, $J$ are suddenly switched to $-\Gamma$,
$-\omega_{0}$, $-J$, respectively, i.e. the Hamiltonian is switched
to ${\cal H}[-\Gamma,-\omega_{0},-J]$ instantaneously. Because
$|\xi_1[\Gamma,\omega_{0},J]\rangle =
|\xi_2[-\Gamma,-\omega_{0},-J]\rangle$,
$e^{i(\gamma_1^B[\omega_{0},J]-E_1[J]\tau)}|\xi_1[\Gamma,\omega_{0},J]\rangle$
is an eigenstate of the switched Hamiltonian ${\cal
H}[-\Gamma,-\omega_{0},-J]$, but with eigenvalue $E_2[-J]$.
Therefore this state continues to be an instantaneous eigenstate in
the second cycle, accumulating a phase
$\gamma_2^B[-\omega_{0},-J]-E_2[-J]\tau$ in the second cycle.
Therefore, the total phase at the end of these two cycles is
$$\gamma_1^B[\omega_{0},J]-E_1[J]\tau +
\gamma_2^B[-\omega_{0},-J]-E_2[-J]\tau= 2\gamma_1^B[\omega_{0},J].
$$ The situation is similar if the initial state is
$|\xi_2[\Gamma,\omega_{0},J]\rangle$, for which the total phase is
$2\gamma_2^B[\omega_{0},J]$.

If the initial state is $|\xi_3[\Gamma,\omega_{0},J]\rangle$, the
situation is the following. It evolves  as an instantaneous
eigenstate of the Hamiltonian in the first cycle, accumulating a
phase $\gamma_3^B[\omega_{0},J]-E_3[J]\tau$. At the end of the first
cycle, $\Gamma$, $\omega_{0}$, $J$ are instantaneously switched to
$-\Gamma$, $-\omega_{0}$, $-J$, respectively. Because
$|\xi_3[\Gamma,\omega_{0},J]\rangle =
|\xi_3[-\Gamma,-\omega_{0},-J]\rangle$,
$e^{i(\gamma_3^B[\omega_{0},J]-E_3[J]\tau)}|\xi_3[\Gamma,\omega_{0},J]\rangle$
is also an eigenstate of the switched Hamiltonian ${\cal
H}[-\Gamma,-\omega_{0},-J]$, but with eigenvalue $E_3[-J]$. Thus
this state continues to be an instantaneous eigenstate of the
switched Hamiltonian in the second cycle, accumulating a phase
$\gamma_3^B[-\omega_{0},-J]-E_3[-J]\tau$. Therefore, the total phase
accumulated in these two cycles are
$$\gamma_3^B[\omega_{0},J]-E_3[J]\tau+
\gamma_3^B[-\omega_{0},-J]-E_3[-J]\tau= 2\gamma_3^B[\omega_{0},J].
$$ The situation is similar if the initial state is
$|\xi_4[\Gamma,\omega_{0},J]\rangle$, for which the total phase in
the two cycles is $0$.

These four paths, each starting with an instantaneous eigenstate of
the Hamiltonian, in the two cycles with opposite signs of $\Gamma$,
$\omega_{0}$ and $J$,  can be summarized  as
\begin{equation}
\begin{array}{l}
|\xi_1\rangle \rightarrow
e^{i\gamma_1[\Gamma,\omega_{0},J]}\xi_1\rangle \rightarrow
e^{2i\gamma_1^B[\Gamma,\omega_{0},J]}|\xi_1\rangle,\\
|\xi_2\rangle \rightarrow
e^{i\gamma_2[\Gamma,\omega_{0},J]}\xi_2\rangle
\rightarrow e^{2i\gamma_2^B[\Gamma,\omega_{0},J]}|\xi_2\rangle,\\
|\xi_3\rangle \rightarrow
e^{i\gamma_3[\Gamma,\omega_{0},J]}\xi_3\rangle \rightarrow
e^{2i\gamma_3^B[\Gamma,\omega_{0},J]}|\xi_3\rangle ,\\
|\xi_4\rangle \rightarrow
e^{i\gamma_4[\Gamma,\omega_{0},J]}\xi_4\rangle \rightarrow
|\xi_4\rangle,
\end{array} \label{xi}
\end{equation}
where each arrow represents a cycle.

In this way we realize a purely geometric unitary transformation, as
described in Eq.~(\ref{geo}),  with
\begin{equation} U_B=diag(e^{2i\gamma_1^B}, e^{2i\gamma_2^B},
e^{2i\gamma_3^B},1).\end{equation}   $U_B$ may be used as a
two-qubit phase gate, if the logic qubit basis states are encoded as
the eigenstates $|\xi_n\rangle$'s.

A question may arise why all the signs of $\Gamma$, $\omega_0$ and
$J$ need to be reversed in the second cycle, given that neither
$E_n$ nor $\gamma_n^B$ depend on the sign of $\Gamma$, and that
$E_n$ does not depend  on the sign of $\omega_0$ either. The crucial
reason is to establish a one-to-one correspondence between each
instantaneous eigenstate in the first cycle and an instantaneous
eigenstate in the second cycle, so that if the initial state is an
instantaneous eigenstate, it can adiabatically evolve as an
instantaneous  eigenstate throughout the two cycles. Indeed, as
implied by the expression of the eigenstates in Eq.~(\ref{xin}),
only if signs of all the three quantities are reversed, can we
establish the equality of each eigenstate of the original
Hamiltonian and  an eigenstate of the switched Hamiltonian, as given
in (\ref{sr}). Happily, the corresponding eigenvalues of the two
Hamiltonians have a same magnitude and opposite signs, as given in
(\ref{berry}). Such correspondence does not exist if all the signs
of the three quantities are not reversed. For example, if only the
signs of $\omega_0$ and $J$ are reversed, an eigenstate of the
original Hamiltonian has nothing to do with any eigenstate of the
switched Hamiltonian, although $|\xi_n[\Gamma,\omega_0,J]\rangle$
and $|\xi_n[\Gamma,-\omega_0,-J]\rangle$ have equal Berry phases and
eigenvalues. Therefore, an eigenstate in the first cycle does not
evolve to an eigenstate in the second cycle with signs of $\omega_0$
and $J$ reversed, consequently, starting as any instantaneous
eigenstate, the evolution is not a matter of a simple phase factor
in the whole process.

One might wonder why the sign reversal of the $\Gamma$  is
necessary, as $xy$-component of the field  rotates. To avoid this
confusion, one should note that the sign reversal of $xy$-component
should be done together with the sign reversal of $z$-component of
the field as well as $J$, such that an instantaneous eigenstate in
the first continues to be an instantaneous eigenstate in the second
cycle.

Additional robustness comes from the situation that  the
instantaneous eigenvalues are constant. This implies that not only
there is no restriction on how $\theta$ varies with time in each
cycle, but also $\theta$ in each cycle does not need to have any
relation with that in the other cycle.

Reversing the signs of both $\omega_0$ and $\Gamma$ simultaneously
means reversing the direction of the total field. $J$ is also
controllable and can be reversed in some artificial systems such as
cold atoms in optical lattices or superconducting qubits in
Josephson junction circuits. To give an illustrative example
demonstrating the feasibility, consider the superconducting charge
qubits, based on Cooper pair boxes. The sign of the coupling
constant $J$ can be reversed in some coupling schemes for
superconducting charge qubits. For example, in the scheme of Averin
and Bruder~\cite{averin}, two charge qubits are coupled through a
variable electrostatic transformer, with the effective interaction
$J \sigma_{az}\sigma_{bz} + \delta (\sigma_{az}+\sigma_{bz})$, where
$\sigma_{az}$ and $\sigma_{bz}$ are simply related to the numbers
($0$ and $1$) of Cooper pairs in the boxes,  $\delta$ can be set to
be zero by choosing appropriate values of capacitances, and the
coupling constant $J$ can be reversed by making suitable change of
the gate voltage of the transformer. Now, one may additionally also
follow the experiment by Leek {\it et al}, where Berry phase of a
single superconducting charge qubit was observed~\cite{leek}. Each
Cooper pair box, while remaining in the coupling circuit, can be
embedded in a one dimensional microwave transmission line
resonator~\cite{wallraff}, hence coupled to microwave radiation,
realizing the coupling of a qubit with a field with both a static
and a rotating component in the rotating frame. The coupling with
radiation does not affect the simplification of the effective
coupling in Ref.~\cite{averin}, because the system is in the
Born-Oppenheimer regime. Consequently, one implements the total
Hamiltonian ${\cal H}_a+{\cal H}_b+J \sigma_{az}\sigma_{bz}$, where
$H_a$ and $H_b$ are Hamiltonians of single qubits coupled with the
rotating field.  The reversal of the sign of $\omega_0$ can be
achieved by changing the gate change, while the reversal of the sign
of $\Gamma$ can be achieved by adding an extra phase in the
microwave radiation. Note that there may be other implementations
realizing our scheme. In principle there is no reason that it cannot
be done.

\section{Aharonov-Anandan Phases}

\subsection{Exact Evolution of an arbitrary initial state}

In this section,  we discuss  nonadiabatic geometric phases by
exactly solving the Schr\"{o}dinger equation
\begin{equation} i\frac{\partial}{\partial
t} |\psi(t)\rangle = {\cal H} (t)|\psi (t)\rangle, \label{sch}
\end{equation}
where ${\cal H}(t)$ is still as given in Eq.~(\ref{ham}), but there
is no requirement of adiabaticity.

Because ${\cal H}_{\alpha 1}$ can be written as ${\cal H}_{\alpha
1}=\frac{\Gamma_{\alpha }}{2} e^{-i\omega_{1}t\sigma_{\alpha z}/2}
\sigma_{\alpha x} e^{i\omega_{1} t\sigma_{\alpha z}/2}, $ the total
Hamiltonian can be rewritten as
\begin{equation}
{\cal H} (t) = e^{-i\omega_{1}t\sigma_{\alpha z}/2}{\cal H} (0)
e^{i\omega_{1} t\sigma_{\alpha z}/2}, \label{ham2}
\end{equation} where
\begin{equation} {\cal H}(0) =  \frac{1}{2}(\omega_{0}\sigma_{az}
+\omega_{0}\sigma_{bz} + J\sigma_{az}\sigma_{bz} +
\omega_{1}\sigma_{ax} +\omega_{1}\sigma_{bx})\end{equation}  is the
Hamiltonian at $t=0$.

Setting \begin{equation} |\psi(t)\rangle
=e^{-\frac{i}{2}(\omega_{1}\sigma_{az}+\omega_{1}\sigma_{bz})t}
|\tilde{\psi}(t)\rangle,\end{equation}  we obtain
\begin{equation} i\frac{\partial}{\partial
t} |\tilde{\psi}(t)\rangle = \tilde{\cal H} |\tilde{\psi}
(t)\rangle.\end{equation}  where
\begin{widetext} \begin{equation}
\tilde{\cal H} =
 \frac{1}{2}[(\omega_{0}-\omega_{1})\sigma_{az}
+(\omega_{0}-\omega_{1}) \sigma_{bz} + J \sigma_{az}\sigma_{bz} +
\omega_{1}\sigma_{ax} + \omega_{1}\sigma_{bx}] \label{tilde}
\end{equation} \end{widetext}
is time-independent. $\tilde{\cal H}$ is just the Hamiltonian in the
rotating frame. Hence $|\tilde{\psi}(t)\rangle = e^{-i\tilde{H} t}
|\tilde{\psi}(0)\rangle$. Therefore the solution of the
Schr\"{o}dinger equation Eq.~(\ref{sch}), in the original frame, is
given by
\begin{equation} |\psi(t)\rangle =
e^{-\frac{i}{2}(\omega_{1}\sigma_{az}+\omega_{1}\sigma_{bz})t}
e^{-i\tilde{\cal H}t}|\psi(0)\rangle. \label{psit} \end{equation}

\subsection{Cycling States and Aharonov-Anandan Phases}

Like the case of a single spin in a magnetic field~\cite{aharonov},
the state is cyclic in the projected space of rays if starting as an
eigenstate of $\tilde{\cal H}$. After a cycle, the state picks up a
phase factor, which is the sum of a dynamical phase and a geometric
phase called Aharonov-Anandan phase.

Suppose the initial state, in the original frame, is an eigenstate
of $\tilde{\cal H}$, written as
\begin{equation}
|\Psi(0)\rangle=|\tilde{\xi}_n\rangle\equiv\tilde{x}_n|\uparrow
\uparrow\rangle + \tilde{y}_n|\uparrow
\downarrow\rangle+\tilde{z}_n|\downarrow \uparrow\rangle+
\tilde{w}_n|\downarrow \downarrow\rangle, \label{aa0}
\end{equation}
satisfying
\begin{equation} \tilde{\cal H}|\tilde{\xi}_n\rangle=\tilde{E}_n|\tilde{\xi}_n\rangle.\end{equation}
It should be emphasized that though it equals an eigenstate
$|\tilde{\xi}_n\rangle$ of the rotating-frame Hamiltonian,
$|\Psi(0)\rangle$ and the Aharonov-Anandan phase are in the original
frame. At the end of this article, we shall comment on its
difference with Berry phases in the rotating frame.

Then according to Eq.~(\ref{psit}), the state at time $t$ becomes
\begin{widetext}
\begin{equation} |\psi_n(t)\rangle =
e^{-i\tilde{E}_nt}(e^{-i\omega_{1}t}\tilde{x}_n|\uparrow
\uparrow\rangle +\tilde{y}_n|\uparrow
\downarrow\rangle+\tilde{z}_n|\downarrow
\uparrow\rangle+e^{i\omega_{1}t}\tilde{w}_n|\downarrow
\downarrow\rangle), \label{exact}\end{equation}
\end{widetext}
which is always cyclic in the projected space of rays, with period
$\tau=2\pi/\omega_1$, i.e. the period of the Hamiltonian. Because
$\omega_1\tau = 2\pi$,
\begin{equation} |\psi_n(\tau)\rangle =
e^{i\beta_n}|\tilde{\xi}_n(0)\rangle,\end{equation}   where
\begin{equation} \beta_n= -\tilde{E}_n\tau,
\label{gamma}
\end{equation}
is the total phase accumulated in each cycling period.
\begin{equation} \beta_n=\beta^d_n+\beta_n^A ,\end{equation}  where the
dynamical phase is \begin{eqnarray} \beta^d_n & = & -\int_{0}^{\tau}
\langle \psi_n(t)|{\cal H}(t)|\psi_n(t)\rangle dt  \nonumber \\ & =&
-\tilde{E}_n\tau- 2\pi(|\tilde{x}_n|^2-|\tilde{w}_n|^2),
\end{eqnarray} while the Aharonov-Anandan phase $\beta_n^A
\equiv \beta_n-\beta^d_n$ \cite{aharonov} is given by
\begin{equation} \beta_n^A =
2\pi(|\tilde{x}_n|^2-|\tilde{w}_n|^2). \label{a}
\end{equation}

We now come to the explicit form of the eigenstate
$|\tilde{\xi}_n\rangle$ of $\tilde{\cal H}$. Note that
mathematically  $\tilde{\cal H}$ can be obtained from ${\cal H}$ by
replacing $\omega_0$ as $\omega_0-\omega_1$, and setting $\omega_1
t$ as $0$. Hence the eigenstates of $\tilde{\cal H}$ are given by
those of ${\cal H}$ with $\omega_0$ replaced as $\omega_0-\omega_1$,
and $\omega_1 t$ is set to $0$.

One of the eigenstates of $\tilde{\cal H}$ is the singlet state
\begin{equation} |\tilde{\xi}_4\rangle = \frac{1}{\sqrt{2}}(|\uparrow\downarrow\rangle-
|\downarrow\uparrow\rangle),\end{equation}  with energy
$\tilde{E}_4=-J/2$, as can be understood by noting that $\tilde{\cal
H}= (\omega_0-\omega_1) S_z +J S^2_z+\Gamma(S_x\cos \omega_1t +S_y
\sin \omega_1 t)-J/2$, for which the singlet state with total spin
$S=0$ must be an eigenstate. For $n=1,2,3$,
\begin{widetext}
\begin{equation} |\tilde{\xi}_n\rangle=\frac{1}{\sqrt{\tilde{{\cal
N}}_n}} [\frac{-2\Gamma }{2(\omega_{0}-\omega_1)+J-2\tilde{E}_n}
|\uparrow\uparrow\rangle+ |\uparrow\downarrow\rangle+|\downarrow
\uparrow\rangle+ \frac{2\Gamma
}{-2(\omega_{0}-\omega_1)+J-2\tilde{E}_n}|\downarrow
\downarrow\rangle],\end{equation}  where  \begin{equation}
\tilde{{\cal N}}_n= 2+
\frac{4\Gamma^2}{[2(\omega_0-\omega_1)+J-2\tilde{E}_n]^2}+
\frac{4\Gamma^2}{[-2(\omega_0-\omega_1)+J-2\tilde{E}_n]^2}.\end{equation}
\end{widetext}
The corresponding eigenvalues,  satisfying
$(2\tilde{E}_n)^3-J(2\tilde{E}_n)^2-(J^2+4(\omega_0-\omega_1)^2+
4\Gamma^2)(2E_n)+ (J^3-4(\omega_0-\omega_1)^2J+4\Gamma^2J)=0$,  are
given by
\begin{equation} \tilde{E}_n=\frac{1}{2}(\tilde{y}_n+\frac{J}{3}),\end{equation}
where $\tilde{v}_n$'s are solutions of the equation
\begin{equation} \tilde{v}^3+\tilde{p}v+\tilde{q}=0,\end{equation}  with
\begin{equation}
\begin{array}{rcl}
\tilde{p}& = & -[\frac{4J^2}{3} + 4(\omega_0-\omega_1)^2
+4\Gamma^2],\\
q & = & \frac{16}{27}J^3
+\frac{[8\Gamma^2-16(\omega_0-\omega_1)^2]J}{3}.
\end{array}\end{equation}
Therefore
\begin{equation}
\begin{array}{rcl}
\tilde{E}_1 & = & \sqrt{-\frac{\tilde{p}}{3}} \cos\tilde{\Phi} +
\frac{J}{6},\\
E_2 & = & \sqrt{-\frac{\tilde{p}}{3}} \cos
(\tilde{\Phi}+\frac{2\pi}{3}) + \frac{J}{6} , \\
E_3 & = & \sqrt{-\frac{\tilde{p}}{3}} \cos
(\tilde{\Phi}-\frac{2\pi}{3})+ \frac{J}{6} ,
\end{array}\end{equation}
where
\begin{equation} \tilde{\Phi}= \frac{1}{3} \arccos
(-\frac{\tilde{q}}{2\sqrt{-(\frac{\tilde{p}}{3})^3}}).\end{equation}

If the initial state is  $|\tilde{\xi}_n\rangle$, then after $\tau$,
the Aharonov-Anandan phase is
\begin{equation}
\beta_n^A =
\frac{32\Gamma^2(\omega_0-\omega_1)(2\tilde{E}_n-J)}{{\cal
N}_n[(2\tilde{E}_n- J)^2 -4 (\omega_0-\omega_1)^2]^2}\end{equation}
for $n=1,2,3$, while $\beta_4^A=0$.

Note that for any form of interaction, the Aharonov-Anandan phase of
a state starting as an eigenstate of $\tilde{\cal H} \equiv {\cal
H}(0)-\omega_1(\sigma_{az}+\sigma_{bz})/2$  can always be obtained
from the Berry phase of the corresponding  eigenstate  of ${\cal
H}(t)$ by replacing $\omega_0$ as $\omega_0-\omega_1$. But such
correspondence does not exist if $\omega_{a0} \neq \omega_{b0}$ or
$\Gamma_a \neq \Gamma_b$.

\subsection{Canceling Phases}

In Sec.~III on adiabatic case, we discussed how to obtain unitary
transformation  purely based on Berry phases, by canceling the
dynamic phases through two cycles with opposite signs of $\omega_0$,
$\Gamma$ and $J$.

Now one may similarly install  two cycles with opposite signs of
$\omega_0$, $\omega_1$, $\Gamma$ and $J$. Then it is the total
phase, instead of the dynamical phase only, that is canceled. This
is because starting as an eigenstate of $\tilde{\cal H}$,
$-\tilde{E}_n\tau$ is the total phase in a cycle. Reversing the sign
of $\omega_1$ means reversing the direction of rotation of the $xy$
component of the field. In the superconducting charge qubit as
studied in \cite{leek}, the reversal of $\omega_1$ can be realized
by reversing the sign of the phase of the microwave field.

Consequently, after such two cycles, one can realize
\begin{equation} |\psi(2\tau)\rangle = |\psi(0)\rangle \end{equation}
for an arbitrary unknown state. Hence for an arbitrary initial
state, one realizes a cycle with no phase factor, with period
$2\tau$. Therefore, in this way, one cannot construct a non-trivial
two-qubit gates based on Aharonov-Anandan phase.

Besides, for $J\neq 0$, it is impossible to cancel the dynamical
phases only by choosing appropriate parameter values in one cycle,
shown as the following. In order that $\beta_n^d=0$,
$\tilde{E}_\alpha \tau=-\beta_n^A$ has to be satisfied. Then
$\tilde{E}_4=0$, implying $J=0$, which is a contradiction.

\section{Summary and discussions}

To summarize, we have rigorously studied geometric phases, including
both Berry phases and  Aharonov-Anandan phases, of two
Ising-interacting spin-$\frac{1}{2}$'s in a rotating magnetic field.
The two spins are equally  coupled with the field, hence none of the
spins can ever be approximated as decoupled with the $xy$-component
of the field. We calculate the geometric phases of the whole system.

For the adiabatic phases, through two consecutive cycles with
opposite signs of $\omega_0$, $\Gamma$ and $J$, the dynamical phases
in the four instantaneous eigenstates can be canceled
simultaneously, thus one obtains a unitary transformation purely
based on Berry phases. The simultaneous reversal of the signs of
$\omega_0$ and $\Gamma$ is  just the reversal of the total magnetic
field.

We have also studied the exact evolution of the state under the
time-dependent Hamiltonian, without the adiabatic condition. If the
initial state is an eigenstate of the time-independent Hamiltonian
$\tilde{H}$ given in Eq.~(\ref{tilde}), then the state is always
cyclic with a phase factor, which is a sum of the dynamical phase
and the Aharonov-Anandan phase. Through two consecutive cycles with
opposite signs of $\omega_0$, $\omega_1$, $\Gamma$ and $J$, the
total phase factor, rather than the dynamical phase only, is
canceled in each path starting as an eigenstate of $\tilde{H}$.
Consequently, any initial unknown state can evolve back without a
phase factor, i.e. one realizes the cycle in the full Hilbert space,
rather than in the projected space of rays. Reversal of the signs of
these quantities can  be realized in some artificial systems, e.g.
the superconducting qubits.

For two spins with Ising interaction, both the Berry phases and
Aharonov-Anandan phases depend on the interaction constant $J$, in
contrast with the case of isotropic Heisenberg
interaction~\cite{shi1}, which is  conserved in any eigenstate of
${\cal H}(t)$ or $\tilde{H}$. In the present case, the Ising
interaction is not conserved in any eigenstate of ${\cal H}(t)$ or
$\tilde{H}$.

Before finishing this article,  we would like to comment on the
difference between Berry phase in the rotating frame and
Aharonov-Anandan phase. The two phases are different entities though
their expressions may be the same.  Aharonov-Anandan phase is a
phase in the original frame. In order that the state is cyclic with
a phase factor in the original frame, a part of which is
Aharonov-Anandan phase, the initial state in the original frame is
equal to an eigenstate of the rotating-frame Hamiltonian
$\tilde{\cal H}$, i.e. $|\psi(0)\rangle = |\tilde{\xi}_n\rangle$, as
given in Eq.~(\ref{aa0}). The Berry phase in the rotating frame, on
the other hand, is the adiabatic geometric phase achieved in the
rotating frame, for an initial state in the rotating frame being an
eigenstate of the rotating-frame Hamiltonian $\tilde{\cal H}$, i.e.
$|\tilde{\psi}(0)\rangle = |\tilde{\xi}_n\rangle$. The calculation
of Berry phase in rotating frame follows the usual way of
calculating Berry phase, now with Hamiltonian and instantaneous
eigenstates are all those in the rotating frame. The phases observed
in Ref. \cite{jones} and in Ref. \cite{leek} are both Berry phases
in the rotating frame~\cite{wrong}. In Ref.~\cite{shi1}, it was
noted that for two spins with isotropic Heisenberg interaction in a
rotating field, it is impossible to construct a nontrivial two-spin
unitary transformation purely based on Berry phases in the original
frame, or based on Aharonov-Anandan phases when the gyromagnetic
ratio of the two spins are equal. Now we see that the situation for
Berry phases in the rotating frame is similar to that of
Aharonov-Anandan phases, i.e. a nontrivial two-spin unitary
transformation purely based on Berry phases in the rotating frame
can be obtained if and only if the gyromagnetic ratio of the two
spins are different. This explains why $J$-dependent Berry phases
was experimentally observed NMR~\cite{jones}, for which the
interaction is, precisely speaking, isotropic Heisenberg-like.

I thank J. Q. You, S. L. Zhu and P. Zoller for discussions, and E.
Sj\"oqvist for calling my attention to Ref.~\cite{yi} after an
earlier version of the present paper was written. This work is
supported by National Science Foundation of China (Grant No.
10674030), Ministry of Science and Technology of China (Grant No.
2009CB929204) and Shanghai Shuguang Project (Grant No. 07S402).


\begin{thebibliography}{99}
\bibitem{wrong} My comment in \cite{shi1} that what had been
observed in \cite{jones} is Aharonov-Anandan phase is wrong.
\end{thebibliography}
\end{document}